\documentstyle[aps,multicol,epsfig,amssymb]{revtex}
\begin{document}
\title{Clock synchronization with dispersion cancellation} 
\author{V. Giovannetti, S. Lloyd$^*$, L. Maccone, and F. N. C. Wong.}
\address{Massachusetts Institute of Technology, Research
Laboratory of Electronics\\ $^*$Corresponding Author: Department of
Mechanical Engineering MIT 3-160,\\ Cambridge, MA 02139, USA.}
%\date{\today}
\maketitle

\begin{abstract}
The dispersion cancellation feature of pulses which are entangled in
frequency is employed to synchronize clocks of distant parties. The
proposed protocol is insensitive to the pulse distortion caused by
transit through a dispersive medium. Since there is cancellation to
all orders, also the effects of slowly fluctuating dispersive media
are compensated. The experimental setup can be realized with currently
available technology, at least for a proof of principle.
\end{abstract}
\begin{multicols}{2}

If two distant persons want to synchronize their clocks, one of them
(say Alice) sends a pulse, which is bounced back by the other
(Bob). Alice measures the pulse creation and travel time and Bob the
time at which the bounce occurs. By exchanging their measurement
results, each of them may know the time of the other relative to his
own clock. This is the idea underlying the Einstein synchronization
scheme {\cite{einstsyn}}. Of course, if the medium between Alice and
Bob is dispersive, then the pulses they exchange get distorted and the
measurements of the bounce time at Bob's side and of the arrival time
at Alice's side acquire an error, which must be summed to the
intrinsic error in such kind of measurements.  In the actual
technological problem of synchronizing clocks of distant parties
{\cite{nist}}, it seems that the main cause of errors is given by the
(possibly fluctuating) dispersion effects of the medium through which
the pulses travel.  Franson {\cite{franson}} proposed a scheme capable
of suppressing the dispersion effects to all orders, given a suitably
tailored dispersive media. Steinberg, Kwiat and Chiao
{\cite{kwiat1,kwiat2}} proposed and experimentally implemented a
scheme capable of suppressing to first order the effects of dispersion
of arbitrary media. In their scheme the time-resolved coherence
properties of frequency entangled pulses to first order do not acquire
any spread when traveling through a dispersive medium. However their
scheme is not suitable for clock synchronization since the dispersion
cancellation is present only if the time of arrival is {\em not}
determined accurately {\cite{kwiat2,jeff}} and since only the
interferometer arms path length difference is
recovered.

Here a modified version of the Steinberg, Kwiat and Chiao
interferometer is presented, which allows the synchronization of the
clocks of Alice and Bob without being bothered by the pulse distortion
as it travels through the intervening medium. This scheme is an
application of the proposal {\cite{paper}} to employ
frequency-entangled pulses to achieve an accuracy increase in clock
synchronization.  The synchronization protocol employed is quite
different from the Einstein clock synchronization scheme: no time
measurement are needed, and the relative distance or the transit time
between Alice and Bob or any dispersive property of the medium play no
role in the protocol. As in the Einstein protocol, the only (rather
reasonable) hypothesis is that the pulse time of travel is the same
both ways. This allows to also employ the scheme in the presence of
fluctuations in the medium under the requirement that the fluctuations
have a time scale  longer than the pulse travel time.

The setup is realizable with currently available technology, since it
employs conventional parametric down converter crystals as
frequency-entanglement source.

In the first section of this paper the time synchronization protocol
that the proposed scheme employs is described. In
Sect. {\ref{s:setup}} the experimental setup and its features are
presented. In Sect. {\ref{s:math}} the equations of motion of the
system are solved and the hypotheses needed to have dispersion
cancellation are given.

\section{Clock synchronization protocol}\label{s:csp}
In this section, the protocol underlying the proposed experimental
setup is described and compared to the Einstein clock
synchronization. It is a classical protocol {\cite{classical}}: the
quantum mechanical features of the setup that will be introduced in
the following sections are employed only to achieve enhanced accuracy
and dispersion cancellation.

\begin{figure}[hbt]
\begin{center}\epsfxsize=.5
\hsize\leavevmode\epsffile{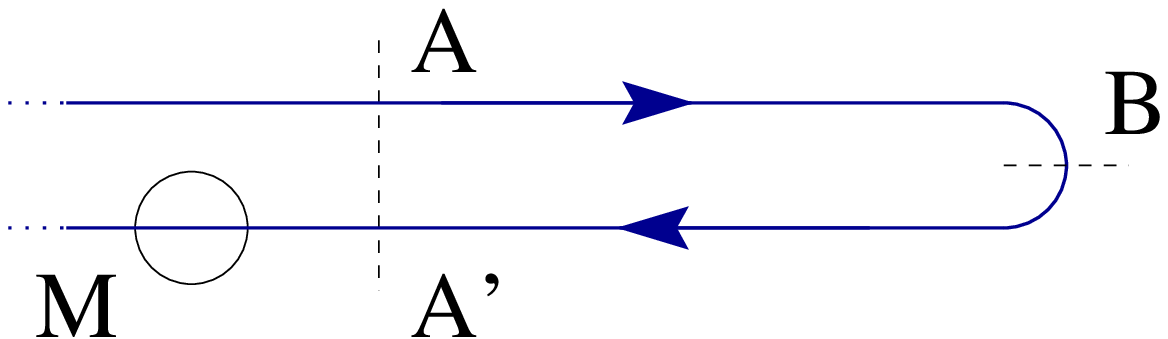} 
\end{center}
\caption{}
\label{f:nastro}\end{figure}

Consider the following scenario for the sake of illustrating the
method: a conveyor belt ({\it i.e.} a physical system in which the
transit time from A to B is the same as from B to A) connects Alice
and Bob (as in Fig. {\ref{f:nastro}}).
Alice pours a quantity of sand which is proportional to the time shown
on her clock on both sides A and A' of the conveyor belt, {\it i.e.},
on the side traveling towards Bob and on the side traveling towards
the point M. Bob, at one end of the conveyor belt (point B), scoops
away a quantity of sand proportional to twice the time shown on his
clock. If the proportionality constant of Alice and Bob are the same,
then the quantity of sand at point M will be constant in time, after
an initial transient when they start to act on the system.  Thus, it
is sufficient to measure such a quantity in order to recover the exact
time difference between Alice's and Bob's clock. The precision with
which this time difference may be recovered depends only on the
precision with which it is possible to measure the quantity of sand at
M.

The main advantage of this scheme over the Einstein clock
synchronization procedure is that no time of arrival measurements need
to be performed. In fact, the time of arrival measurement of a pulse
has an intrinsic unavoidable error dependent on its bandwidth. The
conveyor belt synchronization strikingly allows time synchronization
without any timing measurements. On the other hand, while Einstein's
protocol measures the distance between Alice and Bob (if the pulse
speed is known), this procedure does not allow the recovery of any
information on this distance, unless Bob stops scooping the sand away.

A practical realization of this classical scheme is readily
implemented.  An intense continuous polarized beam which travels from
Alice to Bob and back is employed as conveyor belt. Alice, on her
side, rotates at positions A and A' the beam polarization by an amount
proportional to the time shown on her clock. Bob rotates the
polarization at position B by an amount proportional to twice his
time, but with opposite direction. The shift in polarization measured
at position M allows Alice and Bob to determine the difference in the
time of their clocks up to a rotation period. In the next section, a
scheme is introduced that exploits quantum properties to enhance the
accuracy and cancel the effects of dispersion.

\section{Experimental setup}\label{s:setup}
\begin{figure}[hbt]
\begin{center}\epsfxsize=1.
\hsize\leavevmode\epsffile{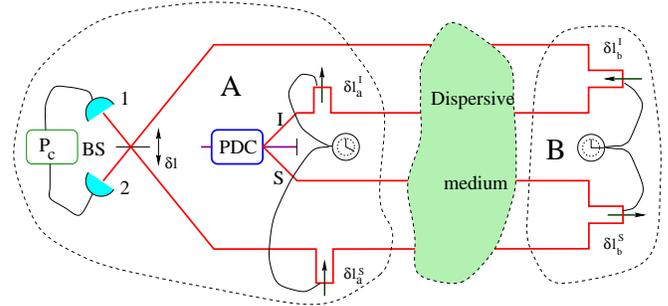} 
\end{center}
\caption{Sketch of the experimental setup for the all-orders
dispersion cancellation synchronization. Alice on the left and Bob on
the right are separated by a dispersive medium. The frequency
entangled state is produced by a parametric down converter crystal
(PDC) and is made to interfere at the 50-50 beam splitter (BS). The
coincidence rate $P_c$ is measured as a function of the position of
the beam splitter $\delta l$. Alice and Bob introduce timing
information in the setup through the time varying delays $\delta
l_a^I,\;\delta l_b^I,\;\delta l_a^S,$ and $\delta l_b^S$. Notice that
two delays ($\delta l_a^I$ and $\delta l_b^S$) increase in time, while
the other two ($\delta l_a^S$ and $\delta l_b^I$) decrease.}
\label{f:experiment}\end{figure}

In this section, an experimental scheme and procedure to implement the
protocol described in the previous section is described.  The
experimental setup is sketched in Fig. {\ref{f:experiment}}. It is
based on the Hong, Ou, and Mandel (HOM) interferometer {\cite{manou}},
which uses the frequency-entangled output beam generated by a
parametric down converter crystal. After propagating through different
optical paths, the signal ($S$) and idler ($I$) beams are interfered
at a beam splitter. By measuring the photon coincidence rate $P_c$ at
the output ports 1 and 2 of the beam splitter, one may acquire very
precise information on the path length difference in the two arms. The
precision limit is given by the inverse of the bandwidth of the
down-converted state (twin beams) and is, to a large extent,
independent on the precision of the photon time of arrival
measurement. Steinberg, Kwiat, and Chiao {\cite{kwiat1,kwiat2}} showed
that in the presence of dispersive media in one arm of the
interferometer, it is possible to cancel to first order the effects of
such dispersion using detectors with a wide integration window. Here
this idea is pursued further. It will be shown that, in the setup
proposed in this paper, the dispersion effects may be canceled to all
orders even if the same dispersive medium is present in both arms of
the interferometer. By using the scheme proposed here, Alice and Bob
may check whether their clocks are synchronized, and may keep them
synchronized through a feedback loop under the hypothesis that the
drift of their clocks is sufficiently slow.

As shown in Fig. {\ref{f:experiment}}, Alice operates the parametric
down converter and generates the twin beams she sends through the
interferometer. The beams travel up to Bob's position and are
reflected back to Alice. She makes them interfere at a 50-50 beam
splitter and measures the coincidence rate of the photodetector clicks
at the outputs of the beam splitter, as a function of the optical path
length difference $\delta l$. She obtains a flat coincidence rate with
a very narrow dip for $\delta l=0$ (as in Fig. {\ref{f:dip}}).

As will be shown in Sect {\ref{s:math}}, in order to check whether
their clocks are synchronized, Alice and Bob must introduce time
varying delays $\delta l^S_a(t)$, $\delta l^I_a(t)$, $\delta
l^S_b(t)$, and $\delta l^I_b(t)$ both on the idler $I$ and signal $S$
beam. (As examples of varying delays consider moving mirrors or time
varying phase shifters). These delays play the role of adding or
subtracting the sand from the conveyor belt. In order to have
dispersion cancellation, four delays are needed instead of only the
two that are sufficient for the synchronization. The protocol requires
these delays to be linked to the time measured by Alice and Bob's
clocks respectively. Consider, for example, the case of linear time
dependence, {\it i.e.}
\begin{eqnarray}
\begin{array}{ll}
\delta l^I_a(t)=v(t-t_0^a)\qquad &
\delta l^I_b(t)=-v(t-t_0^b)\\\\
\delta l^S_a(t)=-v(t-t_0^a)\qquad &
\delta l^S_b(t)=v(t-t_0^b)\;,\end{array}
\;\label{delayt}
\end{eqnarray}
where $t_0^a$ and $t_0^b$ are the times at which Alice and Bob start
their clocks and $v$ is the delay rate ({\it e.g.} the speed of the
moving mirrors). As shown in Fig. {\ref{f:experiment}}, Alice places
one of her delays at the beginning and the other one at the end of the
transmission line.  Bob applies both his delays in the middle of the
two paths. We might assume that Alice and Bob have identical clocks
and only want to find out the time difference $\tau= t_0^b-t_0^a$
between them.  Because of the introduction of the delays
(\ref{delayt}), the final path length difference as measured by the
HOM interferometer will be affected by a factor dependent on $\tau$
and Alice can measure it by observing the shift of the dip position in
the photon coincidence rate $P_c$ as shown in Fig. {\ref{f:dip}}.

\section{Analysis of the experiment}\label{s:math}
In this section the main result is derived. It will be shown that the
dip in the coincidence rate graph is located at the position $\delta
l_{0}=4v\tau$, which shows that by measuring the Mandel dip position
{\cite{manou}}, one can recover the time difference between Alice and
Bob's clock.  There is no dependence on the distance between Alice and
Bob or on any property of the intervening medium. The only two
hypotheses that are required are {\it a)} the travel time of the twin
beam state from Alice to Bob is the same as from Bob to Alice (which
gives a lower time limit for the fluctuations of the intervening
medium) and {\it b)} the medium acts in the same way on the beams
traveling in both directions (which gives a lower limit for the
spatial inhomogeneities of the medium: the distance between the beams
traveling in the two directions must be smaller than the inhomogeneity
characteristic length).

The twin beam state at the output of the crystal (cw pumped at
frequency $2\omega_0$) is given by the maximally frequency entangled
state \begin{eqnarray} |\psi\rangle=\int d\omega
\;\phi(\omega)\;|\omega_0+\omega\rangle_I|\omega_0-\omega\rangle_S
\;\label{state},
\end{eqnarray}
where $|\omega\rangle_S$ and $|\omega\rangle_I$ denote the signal and
idler `frequency states' ({\it i.e.} state in which there is only one
photon at frequency $\omega$ and the vacuum for all the other
frequencies), and $\phi(\omega)$ is the spectral function of the
down-converted light, centered in $\omega=0$ and characterized by the
bandwidth $\Delta\omega$. The coincidence rate at the photodetectors
is given by the Mandel formula for photodetection {\cite{mandel}}
\begin{eqnarray} &&P_c\propto
\int_T dt_1dt_2\;\langle\psi|E_1^{(-)}E_2^{(-)}
E_2^{(+)}E_1^{(+)}|\psi\rangle
\;,\label{mand}
\end{eqnarray}
where $T$ is the integration time window of the detectors. In
Eq. (\ref{mand}) the electromagnetic fields at time $t_j$ at the
output of the beam splitter, assuming for the sake of simplicity a
linear polarization, are given by
\begin{eqnarray} &&
\left\{\begin{array}{l}E_j^{(+)}=i\int
d\omega\;\sqrt{\frac{\hbar\omega}{4\pi cA}}\;a_j(\omega)
e^{-i\omega(t_j-x_j/c)}\cr
E_j^{(-)}=\left(E_j^{(+)}\right)^\dag\qquad\qquad \mbox{for }j=1,2\;,
\end{array}
\right.
\label{campoe}
\end{eqnarray}
where $A$ is the beam cross section and $x_j$ is the position of the
detector. It is possible to match the output fields of the beam
splitter with the fields before the moving mirrors by applying the
beam splitter transformation on the mirrors' reflected fields. This
latter may be obtained by performing some Lorentz transformations on
the input fields: first from the source to the moving-mirror frame and
then back to the laboratory frame.  This procedure yields the
transformations for the field annihilation operators when the fields
are bounced off the moving mirrors.  Analyze the idler beam first. One
finds that the annihilation operator $a_I(\omega)$ at the crystal
position is evolved into $a'_I(\omega)$ at Bob's position (at a
distance $L$ from the crystal) and into $a''_I(\omega)$ at the beam
splitter position (at a distance $L'$ from Bob), with
%Lorenzo: vedi i conti sui fogli M5-M10.
\begin{eqnarray}
a'_I(\omega)&=&\sqrt{\chi}\;a_I(\chi\omega)\;
e^{{-i\omega[\frac{2\beta}{1-\beta}(t_0^a-{x_0}/c)-L/c]+
i\kappa^I_t(\omega)}}\nonumber\\
 a''_I(\omega)&=&a_I(\omega)\times\label{dopplidler}\\&&
e^{{-i\omega[\frac{2\beta}{1+\beta}(t_0^a-t_0^b-
\frac{x_0}c)-\frac {L/\chi+ L'}c]+
i\kappa^I_t(\omega/\chi)+i\kappa^I_f(\omega)}}\nonumber
\;, 
\end{eqnarray}
where $\beta=\frac vc$, $\chi = \frac{1+\beta}{1-\beta}$ is associated
to the Doppler shift introduced by the moving mirrors, and $x_0$ is
the distance of Alice's delay from the crystal. In
Eq. (\ref{dopplidler}) the terms $\kappa^I_t$ and $\kappa^I_f$ take
into account the effect of the dispersive medium on the idler beam on
their way {\it to} and {\it from} Bob respectively. Notice that
because of the Doppler shift introduced by the first mirror (which,
for $\beta>0$, reduces the frequency of the beam after the mirror),
$\kappa^I_t$ is evaluated at $\omega/\chi$. On the other hand, because
of the compensation due to the second mirror, $\kappa^I_f$ is
evaluated at a frequency $\omega$.  Analogous procedure applies to the
signal beam, resulting in the operator transformations
\begin{eqnarray}
a'_S(\omega)&=&\sqrt{\chi}\; a_S(\chi\omega)
\;e^{{-i\omega[\frac{2\beta}{1-\beta}(t_0^b-L/c)- L/c]
+i\kappa^S_t(\chi\omega)}}\nonumber\\
a''_S(\omega)&=&a_S(\omega)\times\label{dopplsign}\\
&&e^{{i\omega[\frac{2\beta}{1+\beta}(t_0^a-t_0^b+
\frac{x_0}c)+\frac {L/\chi+ L'}c]+
i\kappa^S_t(\omega)+i\kappa^S_f(\omega/\chi)}}\;,\nonumber
\end{eqnarray}
where here $a'_S$ has been evaluated after the delay $\delta l_b^S$.

\begin{figure}[hbt]
\begin{center}\epsfxsize=.6
\hsize\leavevmode\epsffile{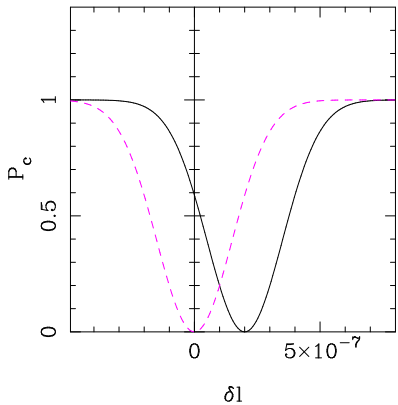} 
\end{center}
\caption{Plot of $P_c$ {\it vs.} $\delta l$. The dashed line
refers to the case $v=0$, the continuous line to the case $v=50\frac
ms$. In this graph, the Gaussian spectrum of the down-converted beams
has a bandwidth $\Delta\omega=10^{15}s^{-1}$ and $\tau=1\ ns$.}
\label{f:dip}\end{figure}

As shown in Fig. {\ref{f:experiment}}, the modes which are detected at
the output of the beam splitter are obtained as\begin{eqnarray}
\left\{\begin{array}{l}a_1(\omega_1)=\frac 1{\sqrt{2}}
\left[ia''_I(\omega_1)\;e^{-i\omega_1\delta l/c}
+a''_S(\omega_1)\right]
\\\\a_2(\omega_2)=\frac 1{\sqrt{2}}
\left[ia''_S(\omega_2)+a''_I(\omega_2)\;
e^{-i\omega_2\delta l/c}\right]
\;,\end{array}\right.\;\label{campibs}
\end{eqnarray}
where $\delta l$ is the delay that Alice introduces in order to scan
the path length dependence of $P_c$. Replacing Eq.  (\ref{state}) and
(\ref{campibs}) into (\ref{mand}), and taking the limit $T\to\infty$,
which corresponds to wide time window at the detection
{\cite{kwiat2,jeff}}, one obtains
\begin{eqnarray} P_c\propto\int d\omega_1d\omega_2|\langle
0|a_1(\omega_1)a_2(\omega_2)|\psi\rangle|^2
\;\label{coinrate},
\end{eqnarray}
where the usual limit $\Delta\omega\ll 2\omega_0$ was employed to
simplify the frequency dependence of the field (\ref{campoe}).
Assuming the symmetry $\phi(\omega)=\phi(-\omega)$ of the spectral
function, the matrix element in (\ref{coinrate}) is given by
\begin{eqnarray} &&\langle 0|a_1(\omega_1)a_2(\omega_2)|\psi\rangle=
\frac 12\;\delta(\omega_1+\omega_2-2\omega_0)
\;e^{i\varphi}
\times
\nonumber\\&& 
\phi({\omega_1}-\omega_0)
\Big[1-e^{{2i({\omega_1-\omega_0})
[\frac{4\beta}{1+\beta}(t_0^b-t_0^a)-\delta 
l/c]-i\Delta\kappa(\omega_1)}}\Big]
\label{panino}\;,
\end{eqnarray}
where $\varphi$ is an overall phase term that will disappear taking
the modulus, and \begin{eqnarray}
&&\Delta\kappa(\omega)=\kappa^S_t(\omega)-\kappa^I_f(\omega)+
\kappa_f^I(2\omega_0-\omega) -\kappa_t^S(2\omega_0-\omega)  
\nonumber\\&&+\kappa_t^I(\frac{2\omega_0-\omega}{\chi})
-\kappa_f^S(\frac{2\omega_0-\omega}{\chi}) +
\kappa^S_f(\frac\omega\chi)-\kappa^I_t(\frac\omega\chi)
\;\label{deltak}
\end{eqnarray}
is the contribution of the dispersion terms. It is immediate to see
that $\Delta\kappa$ vanishes altogether at all orders by requiring
that $\kappa_t^S=\kappa_f^I$ and $\kappa_f^S=\kappa_t^I$, which can be
satisfied by allowing the {\it from} idler beam to propagate at a
distance less than the spatial inhomogeities of the medium from the
{\it to} signal beam and, equivalently, by allowing the {\it to} idler
beam to propagate near the {\it from} signal beam. In this case any
effect of the medium will be erased.  Replacing Eq. (\ref{panino})
into (\ref{coinrate}), one obtains
\begin{eqnarray} 
P_c=\int d\omega\;|\phi(\omega)|^2\;\left[1-\cos\left(2\frac\omega
c\left(\delta l_0-\delta l\right)\right)\right],
\;\label{fin}
\end{eqnarray}
where $\delta l_0 = \frac{4v\tau}{1+\beta}$ or $\delta l_0\simeq
4v\tau$ in the non-relativistic regime.  Finally, in the simple case
of a Gaussian spectrum $|\phi(\omega)|^2$ with variance
$\Delta\omega^2$, Eq. (\ref{fin}) becomes
\begin{eqnarray} &&P_c= 1-e^{-2\Delta\omega^2\;(\delta l-\delta
l_{0})^2/c^2}
\;\label{gauss},
\end{eqnarray}
which, as shown in Fig. {\ref{f:dip}}, features a dip of width
$\Delta\omega^{-1}/4$ centered in $\delta l=\delta l_0$. By measuring
the dip position, one can recover the time difference $\tau$ between
Alice and Bob's clocks.

Notice that, if the parameter $v$ is known with an accuracy $\Delta
v$, then, in the non relativistic regime, the error in the
determination of $\tau$ is given by
\begin{eqnarray}
\Delta\tau=\sqrt{\frac 1{(4\Delta\omega\beta)^2}+\left(\frac{\Delta
v}v\right)^2\tau^2}
\;\label{erroretau}\ ,
\end{eqnarray}
where the first term refers to the intrinsic accuracy in the dip
position measurement. Suppose for example that the clocks are
initially synchronized up to $\tau=1\; ns$ and the twin beam bandwidth
is $\Delta\omega=10^{15}\;Hz$ (as in {\cite{kwiat1}}), then to achieve
an accuracy $\Delta\tau\sim 10^{-1}\;ns$ one has to use $v\sim 500\;
m/s$ with $\Delta v\lesssim 50\; m/s$.

In conclusion, the proposed protocol allows Bob to measure the time
difference $\tau$ between his and Alice's clock, without being
affected by the dispersion of the intermediate medium. The accuracy of
the scheme is dependent only on the bandwidth $\Delta\omega$ of the
twin beam and on the delay rate $v$, namely he can recover $\tau$ with
an error $\sim 1/(\Delta\omega\;v/c)$.

This work was funded by ARDA, NRO, and by ARO under a MURI program.

\end{multicols}
\end{document}